\documentclass[twocolumn,showpacs,aps,prd,superscriptaddress,nofootinbib]{revtex4}

\usepackage{graphicx,fancyhdr,lineno,multirow,rotating,calc,amssymb}

\usepackage{epsfig}
\usepackage{epsf}

\def\etaP{\ensuremath{{\eta'}}}
\def\rhoP{\ensuremath{{\rho'}}}

\def\MeV{{\rm MeV}}
\def\nut{\ensuremath{\nu_\tau}}

\def\br{\ensuremath{{\cal B}}}
\def\brP{\ensuremath{{\cal B}'}}

\def\sto#1{_{({#1})}}

\def\beq{\begin{equation}}
\def\eeq{\end{equation}}

\def\beqa{\begin{eqnarray}}
\def\eeqa{\end{eqnarray}}

\def\pppz{{\ensuremath{\pi^+\pi^-\pi^0}}}
\def\pzq{{\ensuremath{\pi^0\pi^0\pi^0}}}

\def\etapppz{\ensuremath{{\eta\to\pppz}}}
\def\etapzq{\ensuremath{{\eta\to\pzq}}}

\def\etaPpppz{\ensuremath{{\etaP\to\pppz}}}

\def\matrel{\ensuremath{{\cal M}}}

\def\babar{\mbox{\slshape B\kern-0.1em{\footnotesize A}\kern-0.1em
    B\kern-0.1em{\footnotesize A\kern-0.2em R}}}

\def\Getarhopi{\ensuremath{g_{\eta\rho\pi}}}
\def\Grhopipi{\ensuremath{g_{\rho\pi\pi}}}
\def\GetaPrhopi{\ensuremath{g_{\etaP\rho\pi}}}

\def\BR#1{{\ensuremath{{\cal B}}(#1)}}

\def\M{}
\def\Z{}

\def\tauM{\ensuremath{\tau\M}}
\def\rhoM{\ensuremath{\rho\M}}
\def\rhoPM{\ensuremath{\rhoP\M}}
\def\piM{\ensuremath{\pi\M}}
\def\piZ{\ensuremath{\pi\Z}}

\def\azG{\ensuremath{{a_0}}}
\def\azGM{\ensuremath{{a_0\M}}}
\def\az{\ensuremath{{a_0(980)}}}
\def\azP{\ensuremath{{a_0(1450)}}}
\def\ao{\ensuremath{{a_1}}}
\def\azM{\ensuremath{{a_0\M(980)}}}
\def\azPM{\ensuremath{{a_0\M(1450)}}}
\def\aoM{\ensuremath{{a_1\M}}}

\def\channel{\ensuremath{\tauM\to\piM\etaP\nu_\tau}}
\def\tauPiEtaNu{\ensuremath{\tauM\to\piM\eta\nu_\tau}}

\def\ket|#1>{\left|#1 \right>}

\def\bra<#1|{\left< #1 \right|}

\def\bracket<#1|#2>{\setbox0=\vbox{\hbox{$#1$$#2$}}\left<#1\kern1pt \vrule  height\ht0\kern2pt #2\right>}

\def\dirmat<#1|#2|#3>{\setbox0=\vbox{\hbox{$#1$$#2$$#3$}}\left<#1\kern1pt \vrule height\ht0\kern1pt#2\kern1pt \vrule height\ht0\kern1pt #3\right>}

\begin{document}

\title{\large An Estimate of the Branching Fraction of \boldmath $\channel$}
\bigskip

\author{S.~Nussinov}
\affiliation{Tel Aviv University, Tel Aviv, 69978, Israel }
\affiliation{Schmid College of Science, Chapman University, 
              Orange, CA 92866, USA}
\author{A.~Soffer}
\affiliation{Tel Aviv University, Tel Aviv, 69978, Israel }

\date{\today}

\bigskip
\bigskip

\begin{abstract}
We calculate the expected branching fraction of the
second-class-current decay \channel, motivated by a a recent
experimental upper-limit determination of this quantity.
The largest contribution to the branching fraction is due to the
intermediate $\azM$ scalar meson, assuming it is a $\bar ud$ state. 
Smaller contributions arise from $\azPM$, $\rhoM(770)$, and
$\rhoM(1450)$. 
Our calculated values are substantially below the 
experimental upper limit, and are smaller still
if the $\az$ is a four-quark state, as often 
suggested. 
Thus, a precise measurement or tight upper limit has the potential to 
determine the nature of the $\azM$, as well as search for new 
scalar interactions. 

\end{abstract}

\maketitle

\bigskip
\bigskip

\section{Introduction}
\label{sec:introduction}
In a recent paper~\cite{Nussinov:2008gx}, we considered the 
branching fraction of the isospin-violating decay $\tauPiEtaNu$.
We found an expected branching fraction of 
\beq
\br \equiv \BR{\tauPiEtaNu} = (0.3-1.0) \times 10^{-5},
\eeq
in rough agreement with a detailed chiral-perturbation-theory 
calculation~\cite{ref:xpt} and other evaluations~\cite{ref:oldpred}, 
which yielded central values in the range 
\beq
\br = (1.2-1.6) \times 10^{-5}.
\label{eq:xpt}
\eeq
The experimental bound on this branching fraction, $\br < 1.4\times
10^{-4}$~\cite{Bartelt:1996iv}, was obtained by CLEO with an
$e^+e^-$-collision data sample of 
$3.5~{\rm fb}^{-1}$, a fraction of a
percent of currently available integrated luminosity.
The only related high-luminosity measurement is a stringent \babar\
upper limit on the branching fraction of
\channel~\cite{Aubert:2008nj},
\beq
\brP \equiv \BR{\channel} < 7.2 \times 10^{-6} \ \ @ 90\% \ {\rm CL},
\label{eq:babar-limit}
\eeq
obtained with an integrated luminosity of $384~fb^{-1}$.

The fact that the experimental limit is lower than the results summarized in
Eq.~(\ref{eq:xpt}) raises the question of a possible discrepancy between 
theory and experiment. 
Therefore, our goal in this article is to calculate the expected value
of \brP\ and compare it to the experimental limit.
We adapt the methods used in Ref.~\cite{Nussinov:2008gx} to the
present case, noting that a chiral-perturbation-theory 
calculation of this process, as
performed for \tauPiEtaNu\ by Neufeld and
Rupertsberger~\cite{ref:xpt}, would be very useful.

First, we note several similarities and differences
between the calculations of \brP\ and \br: 
\begin{itemize}
\item  
The $\bar uu + \bar dd$ fraction of the wave function which, unlike
the $\bar ss$ and $gg$ parts, contributes to the decay amplitude, may
be smaller for the $\etaP$. While it appears that the magnitude of the
$\bar ss$ part in relation to that of the light quarks is very similar
for both states, the current estimate of the $gg$ fraction of the wave
function, $Z_{gg}$, is $|Z_{gg}|^2 = 0.3 \pm
0.2$~\cite{Escribano:2008rq}. In our calculations we take $Z_{gg}=0$,
as this yields the most conservative limits on \brP, and since the
modification for finite values of $Z_{gg}$ is straightforward.
\item
Calculations of \br\ in Refs.~\cite{Nussinov:2008gx,ref:xpt,ref:oldpred} 
rely on extrapolations utilizing intermediate,
low-mass $J^{PC}=1^{--}$ and $0^{++}$ hadrons.  Obvious intermediate 
states for the decay \tauPiEtaNu\ are the ground-state mesons
$\rho(770)$ and $\az$. In the case of \channel, these are
off-shell processes, and the contributions of these resonances are
suppressed.  On the other hand, we do have now on-shell decays
involving the next $1^{--}$ and $0^{++}$ states. These are the $\rhoP
\equiv \rho(1450)$ and $\azP$, which contribute to
the $P$- and $S$-wave components of the decay, respectively.
\item
The $\rho$ and $\rhoP$ vectors are the quark-model $\bar ud$, $S$-wave
$1^{--}$ ground state and first radial excitation, respectively.
However, the theoretical assignment of the 
\az\ (and, consequently, that of the \azP\ as well) is 
ambiguous, generating the largest uncertainty in both \br\ and 
\brP. Conversely, information on these branching fractions can help
resolve the longstanding dilemma of the ``$\bar KK$-threshold'' state
$\az$.
The significant branching fractions of \az\ and $f_0(980)$ decays to
$\bar KK$, despite the very small phase space, seem inconsistent
with these mesons being the ground states of the quark-model scalar
nonet, motivating a four-quark ($\bar ud \bar ss$)
interpretation~\cite{ref:jaffe}. In this case, the
$\bar ud$ scalar ground state should most likely be identified with
$\azP$. However, this would make the scalar 190~\MeV\ heavier 
than the axial vector state $\ao(1260)$, implying a pattern of 
$L \cdot S$ splitting different from what is observed in any other 
$L=1$, $\bar qq'$ system.
The more appealing possibility, namely, that the two 980-\MeV\
states are indeed just $\bar ud$ states, may have been partially
resurrected in recent work~\cite{Hooft:2008we}, in which 'tHooft's
$\bar u u \bar d d \bar s s$ six-quark vertex was utilized to admix
the 2- and 4- quark states.
\end{itemize}

The plan of this note is as follows.
As we did in Ref.~\cite{Nussinov:2008gx}, we discuss 
separately our estimates of the $P$- and $S$-wave contributions
to \brP.
In Sec~\ref{sec:vector} we present the more robust results for the
$P$-wave part, calculating upper bounds on the contributions
of the $\rhoM$ and $\rhoPM$ using recently published
experimental data involving \etaP\ and $\tauM$ decays.
In Sec~\ref{sec:scalar} we present the less clear-cut estimate
of the $S$-wave component. This contribution depends 
most strongly on whether the $\az$
is a 4-quark state or the $\bar ud$ ground state. 
In any event, our predictions for $\BR{\channel}$ lie significantly
below the \babar\ limit~\cite{Aubert:2008nj}.
A brief summary and future outlook are given in Sec~\ref{sec:discussion}.

\section{The $L=1$ Contribution}
\label{sec:vector}

In Ref.~\cite{Nussinov:2008gx}, we obtained the $L=1$ contribution to
\br\ assuming that it was dominated by the $\rhoM$, an assumption
justified by the large branching fraction
$\BR{\tauM\to\rhoM\nut}$. We compared this branching fraction to
\br\ using the ratio of coupling constants $\Getarhopi / \Grhopipi$,
where \Grhopipi\ was related to the width of the $\rho$, and 
\Getarhopi\ was obtained by analyzing the Dalitz-plot
distribution of the decay $\etapppz$, taking the scalar contribution
to \etapppz\ from $\BR{\etapzq}$.

This procedure is not directly applicable to
\brP, since there is no experimental information
on the Dalitz-plot distribution of the decay \etaPpppz, nor a measurement
of $\BR{\etaP\to\pzq}$.
Therefore, we make use of the fact that the branching fraction
$\BR{\etaPpppz}$ depends on the coupling constant \GetaPrhopi, under
the conservative assumption that the $\rho^\pm$ states
dominate the decay \etaPpppz. This will yield a conservative upper
bound on \GetaPrhopi, from which we obtain an upper bound on the
$\rhoM$ contribution to \channel.
We discuss the likelihood of this assumption and its implications below.

The differential branching fraction of \etaPpppz\ as a function of the 
Dalitz-plot position is given by
\beq
{d\Gamma_{\etaPpppz} \over \Gamma_\etaP} =
{ (\GetaPrhopi \Grhopipi)^2 \over 384 \sqrt{3} \pi^3 } \, 
  {Q^2 \over m_\etaP \Gamma_\etaP}  
  |\tilde\matrel|^2 dX\, dY,
\label{eq:dGamma}
\eeq
where
\beq
Q \equiv m_\etaP - 3m_\pi
\eeq
is the kinetic energy in the decay, and
\beqa
X &\equiv& {\sqrt{3} \over Q} (T_+ - T_-), \nonumber\\
Y &\equiv& {3 \over Q} T_0 - 1
\label{eq:xy}
\eeqa
are the Dalitz-plot variables, with $T_c$ being the kinetic energy of
the pion with charge $c$.
Assuming $\rho$ dominance, we obtain from Eq.~(15) of Ref.~\cite{Nussinov:2008gx}
the reduced matrix element 
\beq
\tilde\matrel = -2 
 {
    r Y -\frac13 r^2 (Y^2 + X^2)
     \over 1 - \frac23 r Y + \frac13 r^2(\frac13 Y^2 - X^2)
 },
\label{eq:matv1.1}
\eeq
where 
\beq
r = {m_\etaP Q \over m_\rho^2 - {1 \atop 3} m_\etaP^2 - m_\pi^2 
  - i \Gamma_\rho m_\rho} 
                  = 1.6 + 0.7 i .
\eeq
The product $(\GetaPrhopi \Grhopipi)^2$ is then found by integrating
Eq.~(\ref{eq:dGamma}) over the Dalitz plot. In the \etapppz\ case, we
exploited the small value of $r$ to simplify the expression by
expanding in $r$. Due to the $O(1)$ value of $r$ for \etaPpppz, we
resort to numerical integration, which yields
\beq
\int |\tilde\matrel|^2 dX\, dY = 2.4.
\label{eq:reduced-integrals}
\eeq
From this we obtain, using 
$\BR{\etaPpppz} = 3.7\times 10^{-3}$~\cite{Naik:2008tb} 
and $\Grhopipi=6.0$~\cite{Nussinov:2008gx},
\beq
\GetaPrhopi < 0.025.
\label{eq:naive-vector-coupling}
\eeq

As a cross check, we apply the procedure to the decay \etapppz,
obtaining $\Getarhopi < 0.52$. This value is to be compared to the one
obtained from the more precise Dalitz-plot analysis in
Ref.~\cite{Nussinov:2008gx}, $\Getarhopi \approx 0.085$.  The factor
of 6 ratio between the results reflects the fact that the procedure
used here yields but a conservative upper bound, obtained by assuming that 
the decay \etaPpppz\ is dominated by the $\rho^\pm$
resonances. 
This assumption is manifestly false, as the \etaPpppz\ Dalitz-plot
distribution is in much better agreement with a flat distribution than
with that expected from $\rho^\pm$ dominance~\cite{Naik:2008tb}. By
contrast, in Ref.~\cite{Nussinov:2008gx}, the value of \Getarhopi\
obtained from the Dalitz-plot distribution yielded good agreement
between the expected and measured values of $\BR{\etapppz}$.

With this point in mind, we proceed to use the upper bound on
\GetaPrhopi\ to calculate the upper bound on the $\rhoM$ contribution
to $\BR{\channel}$. We do this by relating 
$\BR{\tauM\to\rhoM\sto{\piM\etaP}\nut}$ to
$\BR{\tauM\to\rhoM\sto{\piM\eta}\nut}$ via the ratio of coupling 
constants and phase-space factors
\beq
{\BR{\tauM\to\rhoM\sto{\piM\etaP}\nut} \over \BR{\tauM\to\rhoM\sto{\piM\eta}\nut}}
\approx
    \left({\GetaPrhopi \over \Getarhopi} \right)^2
   {V(\tauM\to\rhoM\sto{\piM\etaP}\nut) \over V(\tauM\to\rhoM\sto{\piM\eta}\nut)},
\eeq
where $\rhoM\sto{\piM\etaP}$ indicates that the \rhoM\ is observed in the 
$\piM\etaP$ final state,
and $V(X)$ is the integral over the Dalitz plot of the three-body
decay $X$. The ratio of
phase-space integrals is $0.06$, with up to $15\%$ variation depending
on whether one uses Blatt-Weisskopf and $s-$dependent widths for the
$\rhoM$ and on the choice of angular distribution.
Using 
$\BR{\tauM\to\rhoM\sto{\piM\eta}\nut} = 3.6\times 10^{-6}$~\cite{Nussinov:2008gx}, 
we obtain
\beq
\BR{\tauM\to\rhoM\sto{\piM\etaP}\nut} < 2\times 10^{-8},
\label{eq:final-rho}
\eeq
more than two orders of magnitude below the \babar\ upper
limit, Eq.~(\ref{eq:babar-limit}).

Next, we evaluate the contribution of the on-shell \rhoPM. One expects that
this state, being a radial excitation and hence having a node in its
wave-function, couples to the ground-state particles $\eta$ and
\piM\ more weakly than the \rhoM.  We hypothesize that this \rhoP\
suppression mechanism works equally strongly for the final states
$\piM\etaP$ and $\piM\piZ$, leading to an equality of the ratios of the
squared matrix elements
\beqa
{\BR{\tauM\to\rhoPM\sto{\piM\etaP}\nut} \over \BR{\tauM\to\rhoM\sto{\piM\etaP}\nut}} \,
{V(\tauM\to\rhoM\sto{\piM\etaP}\nut) \over V(\tauM\to\rhoPM\sto{\piM\etaP}\nut)}
  \nonumber\\
  \approx 
{\BR{\tauM\to\rhoPM\sto{\piM\pi}\nut} \over \BR{\tauM\to\rhoM\sto{\piM\pi}\nut}} \,
{V(\tauM\to\rhoM\sto{\piM\piZ}\nut) \over V(\tauM\to\rhoPM\sto{\piM\piZ}\nut)}.
\eeqa
The relevant phase-space integral ratios are
\beqa
{V(\tauM\to\rhoM\sto{\piM\etaP}\nut) \over V(\tauM\to\rhoPM\sto{\piM\etaP}\nut)}
   &\approx& 0.06,
\nonumber\\[1mm]
{V(\tauM\to\rhoM\sto{\piM\piZ}\nut) \over V(\tauM\to\rhoPM\sto{\piM\piZ}\nut)}
   &\approx& 2.5.
\eeqa
We use the upper bound of Eq.~(\ref{eq:final-rho})
and the central value plus one standard deviation 
of the recent Belle result~\cite{Fujikawa:2008ma}
\beq
\sqrt{\BR{\tauM\to\rhoPM\sto{\piM\pi}\nut} \over \BR{\tauM\to\rhoM\sto{\piM\pi}\nut}}
   = 0.15 \pm 0.05 \, ^{+0.15} _{-0.04}
\label{eq:BellerhoPinTau}
\eeq
to obtain the conservative upper limit
\beq
\BR{\tauM\to\rhoPM\sto{\piM\etaP}\nut} < 8\times 10^{-8}.
\label{eq:final-rhoPrime}
\eeq
We note that this is an upper bound both due to the way we use 
Eq.~(\ref{eq:BellerhoPinTau}) and since Eq.~(\ref{eq:final-rho})
is an upper bound.

\section{The $L=0$ Contribution}
\label{sec:scalar}
Calculating the $L=0$ contributions to $\brP$ is not as
straightforward as the $L=1$ case, where one can make use of the
dominant $\rhoM$ coupling to the leptonic vector current.
Therefore, is is important to evaluate the scalar component
using different methods, as has been done for the \tauPiEtaNu\ 
decay~\cite{Nussinov:2008gx, ref:xpt, ref:oldpred}.
It should be noted that these calculation are performed under the
assumption that the relevant scalar resonances are $\bar ud$
states. The coupling of a 4-quark state to the $\bar ud$ scalar
current is ``Zweig-Rule'' suppressed, making it significantly smaller
than the predictions.

Here we perform a more detailed version of the calculation used in 
Ref.~\cite{Nussinov:2008gx}. We begin with the ratio of branching fractions
\beqa
\kern-4mm
R^\azGM_\aoM &\equiv& {\BR{\tauM\to\azGM\nut} \over \BR{\tauM\to\aoM\nut}}
 \nonumber\\
&=&    {p_\azG \over p_\ao} \times 
       {\left|\dirmat<\azGM|V_{h\mu}|0>
         \dirmat<\nut|J^\mu_l|\tauM>\right|^2
    \over 
       \left|\dirmat<\aoM|A_{h\mu}|0>
  \dirmat<\nut|J^\mu_l|\tauM>\right|^2},
\label{eq:basic-a0a1-ratio}
\eeqa
where $\azG$ stands for either \az\ or \azP,
$\ao$ is the $\ao(1260)$,
$p_X$ is the $\tau$-rest-frame momentum of the products of the
decay $\tauM\to X\nut$, 
$V_{h\mu} \equiv \bar\psi_u(x) \gamma_\mu \psi_d(x)$ 
  is the hadronic vector current,
$A_{h\mu} \equiv \bar\psi_u(x) \gamma_\mu\gamma^5 \psi_d(x)$  
is the hadronic axial vector current, and
$J^\mu_l \equiv \bar\psi_{\nut}(x) \gamma^\mu(1 - \gamma^5) \psi_\tau(x)$  \
is the leptonic current.
The calculation of the leptonic parts of this ratio is well defined,
while all the uncertainty in the hadronic parts comes down to a single
parameter $\xi$, which shall be defined shortly. 
With this in mind, we can take the \azGM\ matrix element to be 
\beq
\dirmat<\azGM|V_{h\mu}|0> = f_0 {q_\mu \over m_\azG} 
\dirmat<\azGM|S_h|0>,
\label{eq:scalar-matrel}
\eeq
where $f_0$ is an isospin-violation suppression factor, and
$S_h \equiv \bar\psi_u(x) \psi_d(x)$    is the scalar current operator.
The weak vector current is conserved up to the 
difference between the $u$- and $d$-quark masses, plus a smaller
electromagnetic part that we neglect. Therefore, 
\beq
\partial^\mu V_{h\mu} \approx (m_d - m_u) S_h.
\eeq
Using this relation in Eq.~(\ref{eq:scalar-matrel}) yields
\beq
f_0 = {m_d - m_u \over m_\azG}.
\label{eq:f0}
\eeq

We use the fact that both the \azGM\
and the $\aoM$ are $P$-wave states to relate the axial and scalar
decay constants
\beq
\dirmat<\aoM|A_\mu|0> = \xi \epsilon_\mu^* 
   \dirmat<\azGM|S|0>.
\label{eq:xi}
\eeq
We note that this is reminiscent of applying $SU(6)$~\cite{ref:su6} or, 
in this case, just $SU(4)$~\cite{Wigner:1936dx} flavor-spin 
symmetry
to the ($L=0$) 15-plet plus singlet 
containing the $\pi$, $\rho$, $\eta$, and $\omega$,
or the ($L=1$) states
$\azG$, $\ao$, $f_0$, and $h_1$.

Naively, one expects $\xi$ in Eq.~(\ref{eq:xi})
to be of order unity. However, 
this parameter incorporates all the hadronic uncertainty in 
our procedure.
With Eqs.~(\ref{eq:scalar-matrel}-\ref{eq:xi}),
Eq.~(\ref{eq:basic-a0a1-ratio}) becomes, after spin averaging and
index contraction,
\beqa
R^\azGM_\aoM  &=& 
  |\xi|^2
  {p_\azG \over p_\ao}
   \left({m_d - m_u \over m_\azG}\right)^2 
  \nonumber\\
&\times&
   {m_\tau^2 - m_\azG^2 \over m_\tau^2 - m_\ao^2}
   \left({m_\ao \over m_\azG}\right)^2
   {1 \over 1 + 2 \left({m_\tau / m_\ao}\right)^2}.
\label{eq:final-ratio}
\eeqa
This yields the branching fractions 
\beqa
\BR{\tauM\to\azM\nut}  &=& 1.6 \times 10^{-6} \, |\xi|^2, \nonumber\\
\BR{\tauM\to\azPM\nut} &=& 6.4 \times 10^{-8} \, |\xi|^2,
\label{eq:raw-scalar-brs}
\eeqa
where, as in Ref.~\cite{Nussinov:2008gx}, we chose the mass difference
of the two light quarks to be $4~\MeV$~\cite{ref:pdg08} and, assuming
that the $\tauM\to 3\pi\nut$ decay is dominated by the \aoM, we took
$\BR{\tauM\to\aoM\nut} = 0.18$. We compare $\BR{\tauM\to\azM\nut}$ of
Eq.~(\ref{eq:raw-scalar-brs}) with the value $\br = 1.2\times
10^{-5}$, obtained from the more elaborate calculation of
Ref.~\cite{ref:xpt}, minus the $\rhoM$ contribution to $\br$, which is
$3.6\times 10^{-6}$~\cite{Nussinov:2008gx}.  This yields
$|\xi|^2\approx 5$, from which we conclude
\beq
\BR{\tauM\to\azPM\nut} \approx 3 \times 10^{-7}. 
\label{eq:corrected-scalar-br}
\eeq
The \azP\ contribution to \channel\ depends also on the branching
fraction $\BR{\azPM\to\piM\etaP}$, regarding which there is only partial
information. However, from the branching-fraction measurements that
have been made~\cite{ref:pdg08}, it is clear that
$\BR{\azPM\to\piM\etaP} < 0.3$.  Hence
\beq
\BR{\tauM\to\azPM\sto{\piM\etaP}\nut} < 1 \times 10^{-7}. 
\label{eq:corrected-scalar-br-to-etaPpi}
\eeq
If the \azPM\ is a radial excitation, which is the case if the \azM\
is the $\bar ud$ ground state, then $\BR{\tauM\to\azPM\sto{\piM\etaP}\nut}$
should be suppressed by an additional wave-function overlap factor.


Next, we look at the contribution of the \azM\ to \channel,
which can be extracted from the relation
\beq
{\BR{\tau\to\nu\az\sto{\pi\etaP}} \over \BR{\tau\to\nu\az\sto{\pi\eta}}}  
  =
  {V(\tau\to\nu\az\sto{\pi\etaP}) \over V(\tau\to\nu\az\sto{\pi\eta})}
R^\etaP_\eta,
\label{eq:a0-ratio}
\eeq
where 
\beq
R^\etaP_\eta \equiv 
\left|{ \matrel(\azM\to\piM\etaP) \over \matrel(\azM\to\piM\eta) }\right|^2
\eeq
is the square of the ratio between the relevant hadronic-decay matrix elements.
We assume that $R^\etaP_\eta$ equals the corresponding ratio 
of \azPM-decay matrix elements, and is hence obtained from
\beq
R^\etaP_\eta 
 \approx 
    {\BR{\azPM\to\piM\etaP}  \over  \BR{\azPM\to\piM\eta}} \times
   {p_\eta  \over  p_\etaP},
\label{eq:a0-matrel-ratio}
\eeq
where $p_X$ is the $\azPM$-rest-frame momentum of the products of the
decay $\azPM\to\piM X$. Given the $\sim 50\%$ error~\cite{ref:pdg08}
on the ratio of branching fractions appearing in
Eq.~(\ref{eq:a0-matrel-ratio}) and the uncertainty on 
the \azPM\ width, $R^\etaP_\eta$ comes out in the range
$[0.25 , 1.25]$.
The ratio of the phase-space integrals in Eq.~(\ref{eq:a0-ratio}) is
0.06, with some dependence on what one takes for the \azM\ width.  
Using the range for \br\ from Eq.~(\ref{eq:xpt}), we
obtain
\beq
\BR{\tau\to\az\sto{\pi\etaP}\nut}   \approx   [0.2 ~{\rm to}~ 1.2] \times 10^{-6}.
\label{eq:final-a0}
\eeq

\section{Conclusions}
\label{sec:discussion}
Combining Eqs.~(\ref{eq:final-rho}), (\ref{eq:final-rhoPrime}),
(\ref{eq:corrected-scalar-br-to-etaPpi}), and~(\ref{eq:final-a0}), we
obtain the branching fraction limit
\beq
\BR{\tau\to\pi\etaP\nut}   <   1.4 \times 10^{-6},
\label{eq:final-summed}
\eeq
in no conflict with the experimental upper limit,
Eq.~(\ref{eq:babar-limit}), which is about five times greater.  Our
result is dominated by the \az\ contribution, assuming it is a $\bar
ud$ state.

The experimental limit was obtained with only a third of the currently
available \babar\ and Belle data sets, and with the $\eta$
reconstructed only in the $\gamma\gamma$ final state. Therefore,
an improvement in  the limit can be expected from the current
generation of $B$ factories, but probably not to the 
level of Eq.~(\ref{eq:final-summed}). By contrast, a Super $B$ 
factory~\cite{Bona:2007qt}, with two orders of magnitude more
luminosity, will be able to use \br\ and \brP\ to investigate the 
nature of the \azM\ and to search for new interactions mediated
by heavy scalars~\cite{Nussinov:2008gx}.

\bigskip
\begin{acknowledgments}
This research was supported in part
by grant number 2006219 from the United States-Israel Binational
Science Foundation (BSF), Jerusalem, Israel. 
The authors thank Leonid Frankfurt and Swagato Banerjee for useful 
suggestions and discussions.
\end{acknowledgments}

\end{document}